\documentclass[12pt]{article}
\usepackage{amsmath}
\usepackage{epsf}
\usepackage{epsfig}
\usepackage{here}
\usepackage{amssymb}
\usepackage{citesort}
\usepackage{graphicx}
\usepackage{latexsym}
\newcommand{\be}{\begin{equation}}
\newcommand{\ee}{\end{equation}}
\newcommand{\bear}{\begin{eqnarray}}
\newcommand{\ear}{\end{eqnarray}}
\newcommand{\e}{\mbox{e}}

\newsavebox{\LSIM}
\sbox{\LSIM}{\raisebox{-1ex}{$\ \stackrel{\textstyle<}{\sim}\ $}}
\newcommand{\lsim}{\usebox{\LSIM}}
\newsavebox{\GSIM}
\sbox{\GSIM}{\raisebox{-1ex}{$\ \stackrel{\textstyle>}{\sim}\ $}}
\newcommand{\gsim}{\usebox{\GSIM}}

\begin{document}
\begin{titlepage}
\begin{flushright}
DESY-03-037 \\
hep-ph/0303183
\end{flushright}
$\mbox{ }$
\vspace{.1cm}
\begin{center}
\vspace{.5cm}
{\bf\Large Flavor violation and warped geometry}\\[.5cm]
Stephan J. Huber\footnote{stephan.huber@desy.de}

\vspace{1cm} {\em 
Deutsches Elektronen-Synchrotron DESY, Hamburg, Germany
}
\end{center}
\bigskip\noindent
\vspace{1.cm}
\begin{abstract}
Extra dimensions have interesting consequences for
flavor physics. We consider a setup where the standard model
fermions and gauge fields reside in the bulk of a warped
extra dimension. Fermion masses and mixings are explained
by flavor dependent fermion locations, without relying on hierarchical 
Yukawa couplings. We discuss various flavor violating processes 
induced by (Kaluza-Klein) gauge boson exchange and 
non-renormalizable operators. Experimental constraints
are satisfied with a Kaluza-Klein scale of about 10 TeV.
Some processes, such as muon-electron conversion, 
are within reach of next generation experiments.
\end{abstract}
\end{titlepage}
\section{Introduction}
Models with extra dimensions have attracted great attention in recent years
as they offer new perspectives
on challenging problems in modern physics. It was demonstrated
by Randall and Sundrum that a small but warped extra dimension 
provides an elegant solution to the gauge hierarchy 
problem \cite{RS} (see also \cite{G}). The fifth dimension 
is an $S_1/Z_2$ orbifold with an AdS$_5$ geometry of curvature $k$, 
bordered by two 3-branes with opposite tensions and separated
by distance $R$.  The red shift induced by the
AdS warp factor $\Omega=e^{-\pi k R}$
generates an exponential hierarchy between the
energy scales on the two branes.
If the brane separation is $kR\simeq 11$, the scale on 
the negative tension brane is of TeV-size, while the scale on 
the other brane is of order $M_{\rm Pl}$.  The 
AdS curvature $k$ and the 5D Planck mass $M_5$ 
are assumed to be of order $M_{\rm Pl}$. Gravity is weak at the 
TeV-brane because the zero mode corresponding to the 4D
graviton is localized at the positive tension brane (Planck-brane). 

In contrast to the original proposal \cite{RS} we take the standard 
model (SM) fermions and gauge bosons as bulk fields. In
the non-supersymmetric framework we are studying, the Higgs field
has to be confined to the TeV-brane in order to preserve the solution to the 
gauge hierarchy problem \cite{CHNOY,HS}. Our motivation is twofold. 
With the SM fermions residing in the 5-dimensional bulk, the 
hierarchy of quark and lepton masses can be related to a higher
dimensional geography \cite{GP,HS2}. Different fermion flavors are localized at 
different positions in the extra dimension or, more precisely, 
have different wave functions. The fermion masses are in direct 
proportion to the overlap of their wave functions with the Higgs field \cite{AS}.
Also the CKM mixing can be explained along these lines. 
A second point is that at the TeV-brane non-renormalizable
operators, now only TeV-scale suppressed, are known to induce 
rapid proton decay, large neutrino masses and flavor violating 
interactions. Because of the warp factor, the effective cut-off 
scale varies along the extra dimension. If the quarks and leptons are 
localized towards the Planck-brane in the extra dimension, the suppression
scales of dangerous operators can be significantly enhanced \cite{GP,HS2}.
Small Majorana neutrino masses can then arise from dimension-five 
interactions, without introducing new degrees of freedom \cite{HS4}. 
The atmospheric and solar neutrino anomalies can be satisfactorily resolved.
Alternatively, Dirac neutrino masses can be generated by a coupling to
right-handed neutrinos in the bulk \cite{GN,HS3}.

Fields living in the 5D bulk can be expanded as a tower of Kaluza-Klein
(KK) states. Electroweak observables, in particular the weak 
mixing angle and gauge boson masses, require the KK
excitations of SM particles to be heavier than about 10 TeV \cite{HS,HLS,HPR}. 
A small hierarchy separates the weak and KK scales. 
In the presence of brane-localized kinetic terms the bound on
the KK scale may be somewhat relaxed \cite{branekin}, but we are not
considering this possibility in the following. 
For fermions confined to the TeV-brane, the electroweak constraints
have been analyzed in ref.~\cite{CET}.

In this paper we address the issue of flavor violation in the warped
SM which is an immediate consequence of our approach to the fermion 
mass problem. Some aspects of this topic have already been discussed in 
the literature \cite{HPR,GP,HS2,K00,dAS,KKS02,B02}. The
crucial connection to the issue of fermion masses, however, has
not been thoroughly investigated so far. This connection allows us to 
obtain more reliable predictions for the flavor violating rates.
Moreover, we are including lepton flavor violating processes which
have not yet been considered in context of the warped SM.
Some results have already been presented in ref.~\cite{H02}.

In the next section we summarize some results on bulk fermions
and gauge bosons in a warped background. In section 3 we are
using the observed fermion masses and mixings to determine the
locations of the various fermion flavors in the extra dimension.
The fermion mass pattern can be accommodated without relying
on hierarchical Yukawa couplings. In section 4 we discuss how the
mixing between different KK levels leads to a non-unitary CKM
matrix. Flavor violation by (KK) gauge boson exchange, being a 
natural consequence of our setup, is studied in section 5.
We find that with a KK scale of 10 TeV the model is consistent
with the experimental constraints, while some processes, 
such as muon-electron conversion, are within reach of 
next generation experiments. In section 6 we show that contributions
from non-renormalizable operators to flavor violating processes
are naturally within experimental bounds, while proton decay
cannot by adequately suppressed. Finally, in section 7 we conclude.

\section{Bulk fields in a warped background}
We assume that the standard model gauge bosons and
fermions reside in the bulk of the warped 5D space-time \cite{RS}
\begin{equation} \label{met}
ds^2=e^{-2\sigma(y)}\eta_{\mu\nu}dx^{\mu}dx^{\nu}+dy^2,
\end{equation}
where $\sigma(y)=k|y|$.  
The 4-dimensional metric is $\eta_{\mu\nu}={\rm diag}(-1,1,1,1)$
and $y$ denotes the fifth coordinate.
The  AdS curvature $k$ is related to the bulk cosmological constant
and the brane tensions.
To set the notation let us briefly review 
some properties of gauge and fermion fields in a slice of AdS$_5$.

A gauge field propagating in a curved background with metric $G^{MN}$ 
is described by the equation of motion
\begin{equation} \label{EL}
\frac{1}{\sqrt{-G}}\partial_M(\sqrt{-G}G^{MN}G^{RS}F_{NS})-M_A^2G^{RS}A_S=0.
\end{equation}
The mass term $M_A^2$ arises from spontaneous symmetry
breaking and is present only for the weak gauge bosons. Since
the Higgs field is localized at the TeV-brane, we have
\begin{equation}\label{a}
M_A^2=\frac{1}{2}(g^{(5)})^2 v_0^2\delta(y-\pi R),
\end{equation}  
where $g^{(5)}$ is the 5D gauge coupling. The Higgs
vev $v_0$ is expected to be of order $M_{\rm Pl}$.
Imposing the gauge $A_5=0$, we decompose the 5D fields as
\begin{equation} \label{KKA}
A_\mu({x},y)=\frac{1}{\sqrt{2\pi R}}\sum^{\infty}_{n=0} A_\mu^{(n)}({x})f_n^A(y).
\end{equation}

Inserting the metric (\ref{met}) and the decomposition (\ref{KKA}) into the
equation of motion (\ref{EL}), the wave functions $f_n^A$
have to satisfy
\begin{equation}
(\partial _y^2-2\sigma'\partial_y-M_A^2+e^{2\sigma}m_n^2)f_n^A=0,
\end{equation}
where $\sigma'=d\sigma/dy$. This equation is solved by \cite{DHRP}
\begin{equation}
f_n^A(y)=\frac{e^{\sigma}}{N_n}\left[J_{1}(\frac{m_n}{k}e^{\sigma})+
                 b^A(m_n)Y_{1}(\frac{m_n}{k}e^{\sigma})\right].
\end{equation}
The spectrum of KK masses $m_n$ and the coefficients $b^A$
are determined by the boundary conditions of the wave functions at the 
branes. The localized mass term $M_A^2$
only affects the boundary condition at the TeV-brane \cite{HS,HLS}. 
The normalization constants are fixed by
\begin{equation}
\frac{1}{2\pi R}\int^{\pi R}_{-\pi R}dy~f_m^A(y)f_n^A(y)=\delta_{mn}.
\end{equation}

In the case $M_A^2=0$ eq.~(\ref{EL}) admits a constant zero
mode solution \cite{DHRP}. The excited states are localized towards the TeV-brane
and have TeV-scale KK masses $m_n\sim n\pi ke^{-\pi kR}$.
In the presence of the mass term $M_A^2$, the zero mode  
acquires a mass 
\begin{equation}\label{mg0}
m_0^2=(g^{(5)})^2v_0^2e^{-2\pi kR}/(2\pi R)+
{\cal O}((g^{(5)})^4v_0^4e^{-4\pi kR}/M_{KK}^2),
\end{equation}
where $M_{KK}=m_1$ denotes the KK scale. The order $1/M_{KK}^2$
corrections are related to a dip in the zero mode wave function
caused by the boundary mass term \cite{HS}. Compared to the 4D standard model
they induce a tree-level shift of the W and Z boson mass ratio relative
to its 4D standard model value.
In refs.~\cite{HS,HLS,HPR} it was shown that the electroweak precision
data thus implies the constraint $M_{KK}\gsim 10$ TeV. 
If the boundary mass term was not included 
in the KK reduction, but evaluated with the $M_A^2=0$ wave functions,
the order $1/M_{KK}^2$ effects would show up 
in the 4D effective action as mixings between the KK states. 

The equation of motion of a fermion in curved space-time reads
\begin{equation}
E_a^M\gamma^a(\partial_M+\omega_M)\Psi+m_{\Psi}\Psi=0,
\end{equation}
where $E_a^M$ is the f\"unfbein, $\gamma^a=(\gamma^{\mu},\gamma^5)$ 
are the Dirac matrices in flat space, and
\begin{equation}
\omega_M=\left(\frac{1}{2}e^{-\sigma}\sigma'\gamma_5\gamma_{\mu},0\right)
\end{equation}
is the spin connection induced by the metric (\ref{met}).
Fermions have two possible transformation properties 
under the $Z_2$ orbifold symmetry,
$\Psi(-y)_{\pm}=\pm \gamma_5 \Psi(y)_{\pm}$. Thus, $\bar\Psi_{\pm}\Psi_{\pm}$ 
is odd under $Z_2$, and the Dirac mass term, which is odd as well, 
can be parametrized as $m_{\Psi}=c\sigma'$. The Dirac mass should 
therefore originate from the coupling to a $Z_2$ odd scalar field 
which acquires a vev. 
On the other hand, $\bar\Psi_{\pm}\Psi_{\mp}$ is even.
Using the metric (\ref{met}) one obtains for the left- and right-handed components
of the Dirac spinor \cite{GN,GP}
\begin{equation}
[e^{2\sigma}\partial_{\mu}\partial^{\mu}+\partial_5^2-\sigma'\partial_5-M_f^2]e^{-2\sigma}\Psi_{L,R}=0,
\end{equation} 
where $M_f^2=c(c\pm1)k^2\mp c\sigma''$ and  $\Psi_{L,R}=\pm\gamma_5\Psi_{L,R}$.

Decomposing the 5D fields as 
\begin{equation}
\Psi(x^{\mu},y)=\frac{1}{\sqrt{2\pi R}}\sum_{n=0}^{\infty}\Psi^{(n)}(x^{\mu})
                  e^{2\sigma}f_n(y),
\end{equation}
one ends up with a zero mode wave function \cite{GN,GP}
\begin{equation}
f_0(y)=\frac{e^{-c\sigma}}{N_0},
\end{equation}
and a tower of KK excited states
\begin{equation}
f_n(y)=\frac{e^{\sigma/2}}{N_n}\left[J_{\alpha}(\frac{m_n}{k}e^{\sigma})+
                 b_{\alpha}(m_n)Y_{\alpha}(\frac{m_n}{k}e^{\sigma})\right].
\end{equation}
The order of the Bessel functions is $\alpha=|c\pm 1/2|$ for $\Psi_{L,R}$.
The spectrum of KK masses $m_n$ and the coefficients $b_{\alpha}$
are determined by the boundary conditions at the 
branes \cite{GN,GP}.
The normalization constants follow from
\begin{equation} \label{normferm}
\frac{1}{2\pi R}\int^{\pi R}_{-\pi R}dy~e^{\sigma}f_m(y)f_n(y)=\delta_{mn}.
\end{equation}

Because of the orbifold symmetry, the zero mode of 
$\Psi_+$ $(\Psi_-)$ is a left-handed (right-handed) Weyl spinor.  
For $c>1/2$ $(c<1/2)$ the fermion is localized near the boundary
at $y=0$ $(y=\pi R)$, i.e.~at the Planck-  (TeV-) brane (see also fig.~\ref{f_1}).

The zero modes of leptons and quarks acquire masses from their 
coupling to the Higgs field
\begin{equation} \label{3.1}
\int d^4x\int dy \sqrt{-G}\lambda^{(5)}_{ij}H \bar\Psi_{i+}\Psi_{j-}
\equiv \int d^4x ~ m_{ij} \bar\Psi_{iL}^{(0)}\Psi_{jR}^{(0)} +\cdots,
\end{equation}
where $\lambda^{(5)}_{ij}$ are the 5D Yukawa couplings. The 
4D Dirac masses are given by
\begin{equation} \label{3.2}
m_{ij}=\int_{-\pi R}^{\pi R}\frac{dy}{2\pi R}\lambda^{(5)}_{ij}H(y) f_{0iL}(y)f_{0jR}(y)
=\frac{l_{ij}v_0}{\pi kR}f_{0iL}(\pi R)f_{0jR}(\pi R).
\end{equation}
In the second step we have used $H(y)=v_0\delta(y-\pi R)/k$
and introduced the dimensionless couplings $l_{ij}=\lambda^{(5)}_{ij}\sqrt{k}$.

The gauge interaction between bulk gauge bosons and fermions,
$g^{(5)}\bar \Psi iE_a^M\gamma^a A_M \Psi$, induces the
effective 4D couplings \cite{GP}
\begin{equation} \label{gc} 
g_{ijn}=\frac{g^{(5)}}{(2\pi R)^{3/2}}\int^{\pi R}_{-\pi R} 
e^{\sigma}f_i(y)f_j(y)f_n^A(y)~dy 
\end{equation} 
between the different KK levels. In the case of a massless 
gauge boson ($f^A_0(y)\equiv1$) this
integral reduces to the normalization condition of fermions
(\ref{normferm}), and one finds $g_{ij0}=\delta_{ij}g^{(5)}/\sqrt{2\pi R}$.
For a massive gauge field the gauge coupling of the zero mode
to TeV-brane fermions is somewhat reduced due to the dip in 
its wave function \cite{HS}. In the SM this effect leads, for instance, 
to smaller gauge couplings of W and Z bosons relative to that of 
the photon. In order to keep these corrections within experimental
bounds we derived in ref.~\cite{HLS} the constraint $M_{KK}\gsim 60$ TeV
if the fermions reside on the TeV-brane. For fermions localized
towards the Planck-brane this constraint becomes weaker than
the 10 TeV bound from the gauge boson masses discussed before.  

Combining eqs.~(\ref{mg0}) and (\ref{gc}), we can determine the Higgs
vev required to provide the measured gauge boson masses. To keep the 
electroweak corrections small enough, we take $M_{KK}=10$ TeV. 
Assuming $k=M_{\rm Pl}$ we find $k R=10.83$ and $v_0=0.043M_{\rm Pl}$.
Thus, there is still a small hierarchy of about twenty between the 
fundamental scale and the Higgs vev.

\section{Quark and lepton masses}
The fermion masses crucially depend on the overlap of the fermion
wave functions and the Higgs profile. They are functions of the 
5D mass parameters of the left- and right-handed fermions, 
$c_L$ and $c_R$ respectively, which enter eq.~(\ref{3.2}).
As the 5D Dirac mass, i.e.~$c$ parameter increases, 
the fermion gets localized closer towards the Planck-brane. 
Its overlap with the Higgs profile at the TeV-brane is reduced, 
which is reflected in a smaller 4D fermion mass from
electroweak symmetry breaking. Since the fermionic zero modes
depend exponentially on the 5D mass parameters, the large hierarchy of
charged fermion masses can be generated from $c$ parameters
of order unity \cite{GP,HS2}. If right-handed neutrinos are introduced in 
the bulk, sub-eV neutrino masses can be explained in the same
manner \cite{GN,HS3}.

Building up the fermion mass matrices from  eq.~(\ref{3.2}) requires
also the specification of the 5D Yukawa couplings.  Thus there are 
considerably more independent parameters in the model than there
are observable fermion masses and mixings. Relating the measured
fermion properties to their locations in the extra dimension needs therefore 
additional assumptions. In ref.~\cite{HS2} it was assumed
that the 5D Yukawa couplings are of order unity, and the fermion mass
pattern is solely due to the different locations.
Additionally, the fermions were localized as closely as possible
towards the Planck-brane in order to maximally suppress the impact of 
non-renormalizable operators on rare processes. 

In the following we are using a somewhat different approach.
While still assuming the 5D Yukawa couplings to be of order unity,
we are looking for a set of $c$ parameters which ``most
naturally'' accounts for the observed fermion masses and mixings.
More precisely, we are taking random  5D Yukawa couplings and
require the averaged fermion properties to fit the experimental data.
Similar methods have been used to study models where the
fermion mass pattern is due to (approximately) conserved charges \cite{NT02}.
There the role of the fermion locations is taken by a set of Higgs
fields.

Let us first focus on the quark sector. At the scale of 10 TeV, where we
are matching our model to the observational data, the (running)
quark masses are 
\begin{eqnarray}\label{qmasses}
&m_u=0.7-2.3~{\rm MeV},\quad & m_c=420-540~{\rm MeV},\quad  m_t=140-148~{\rm GeV}
\nonumber \\
&m_d=1.4-4.2~{\rm MeV},\quad & m_s=28-80~{\rm MeV},\quad m_b=2.1-2.3~{\rm GeV}.
\end{eqnarray} 
We have used one loop renormalization group equations \cite{FK98}
to run the quark masses given in ref.~\cite{FX00} from the scale $M_Z$ to 
10 TeV. The running reduces the quark masses by about 20 percent.
The moduli of the CKM matrix are given by \cite{PDG}
\begin{equation}\label{CKM}
|V_{\rm CKM}|=\left(\begin{array}{ccc} 
0.9741-0.9756 & 0.219-0.226 & 0.0025-0048 \\
0.219-0.226 & 0.9732-0.9748 & 0.038-0.044 \\
0.004-0.014 & 0.037-0.044 & 0.9990-0.9993
\end{array}\right).
\end{equation}
A convention independent measure of CP violation is the
Jarlskog invariant which experimentally is found to be \cite{PDG}
\begin{equation}\label{J}
J=(3\pm0.3)\times 10^{-5}.
\end{equation}

Assuming non-hierarchical 5D Yukawa couplings,
the fermion mass matrix (\ref{3.2}) leads to
a product-like structure
\begin{equation} \label{product}
M\sim\left(\begin{array}{ccc} 
a_1b_1 &a_1b_2  & a_1b_3 \\
a_2b_1 &a_2b_2  & a_2b_3 \\
a_3b_1 &a_3b_2  & a_3b_3 \\
\end{array}\right)
\end{equation}
where $a_i$ and $b_i$ are given by the fermion wave
functions $f_{0iL,R}(\pi R)$. As a function of the fermion location
there is a slow increase $f_0(\pi R)\propto \sqrt{1/2-c}$ 
for $c<1/2$, and an exponential suppression 
$f_0(\pi R)\propto \exp(-c\pi kR)$ for $c>1/2$.
Note that some non-equal Yukawa couplings are needed
to render the mass matrix (\ref{product}) non-singular.
If the mass matrix is diagonalized by $U_LMU_R^{\dagger}$,
the left- and right-handed mixings are typically of order $U_{L,ij}\sim a_i/a_j$
and $U_{R,ij}\sim b_i/b_j$, respectively. Fermions which have
similar positions ($c$ parameters) have large mixings. The mass matrix
(\ref{product}) predicts the approximate relation $U_{13}\sim U_{12}U_{23}$
between the mixing angles, which for the observed CKM matrix (\ref{CKM})
is satisfied up to a factor of about two.

\begin{figure}[t] 
\begin{picture}(100,160)
\put(95,-10){\epsfxsize7cm \epsffile{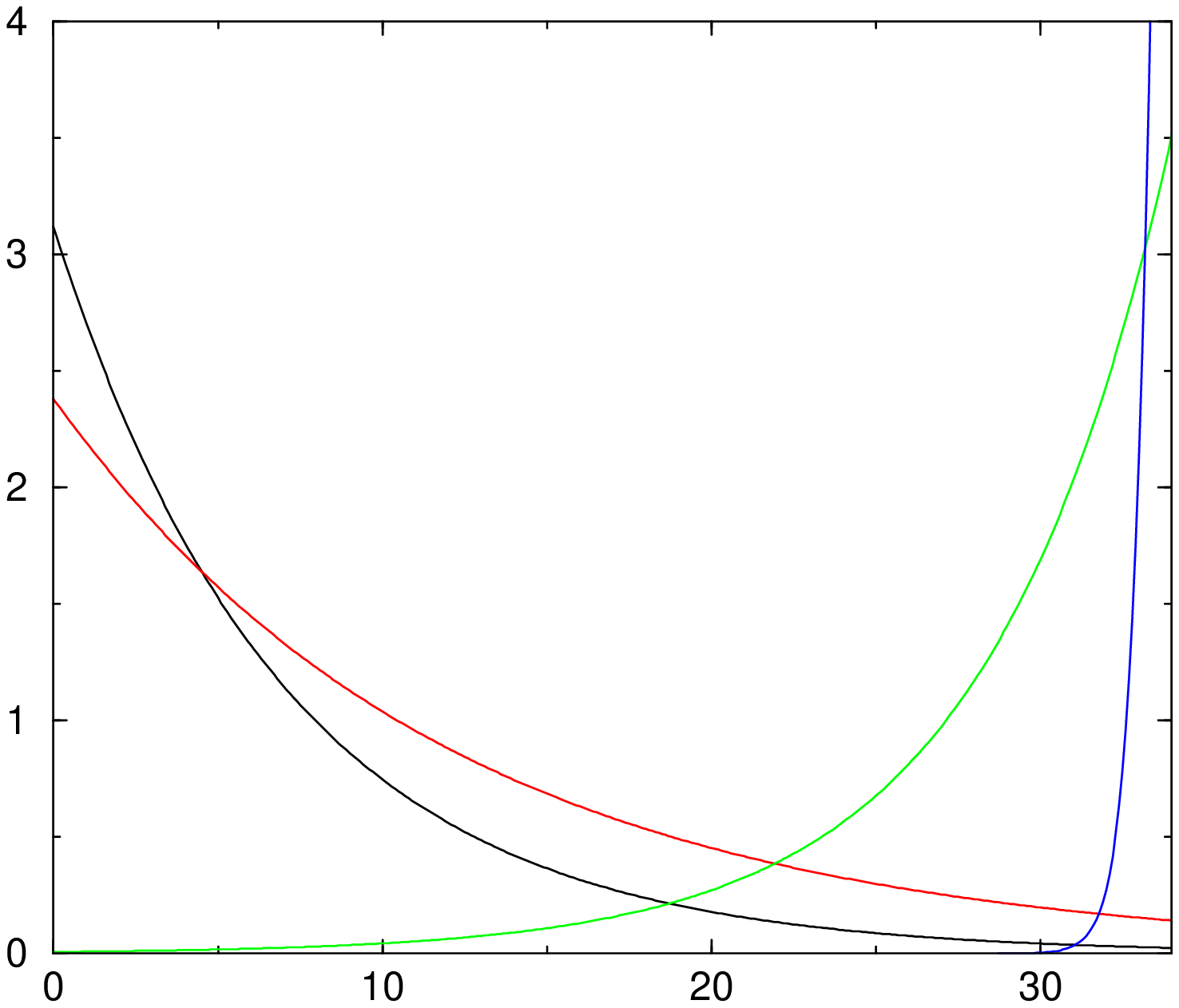}}
\put(45,100){{\large $\e^{\frac{1}{2}\sigma}f_n\uparrow$}} 
\put(120,95){{$q_1^{(0)}$}} 
\put(170,40){{$q_2^{(0)}$}} 
\put(250,55){{$q_3^{(0)}$}} 
\put(270,135){{$q_1^{(1)}$}} 
\put(300,-5){$yk\rightarrow$}
\put(270,165){{TeV-brane}} 
\put(70,165){{Planck-brane}} 
\end{picture} 
\caption{The wave functions of the left-handed quark zero modes $q_i^{(0)}$
and the first KK state, $q_1^{(1)}$, of $Q_1$ 
for the parameters of eq.~(\ref{qlocations}).
}
\label{f_1}
\end{figure}

In our numerical analysis we generate $N=25000$ random sets of up 
and down-type  
Yukawa couplings and diagonalize the emerging mass matrices (\ref{3.2}).
We require the averaged fermion masses and mixings to fit the experimental 
data, using the logarithmic average of a quantity $X$
\begin{equation}
\langle X\rangle =\exp\left(\sum_i^N \frac{\ln(X_i)}{N}\right).
\end{equation}
Taking $2/3<|l_{ij}|<4/3$ and random phases
from 0 to $2 \pi$, we find the ``most natural'' locations
\begin{eqnarray} \label{qlocations}
c_{Q1}=0.643, \quad &c_{D1}=0.643, \quad  &c_{U1}=0.671, \quad  \nonumber \\
c_{Q2}=0.583, \quad &c_{D2}=0.601, \quad &c_{U2}=0.528, \quad\nonumber \\
c_{Q3}=0.317, \quad &c_{D3}=0.601, \quad &c_{U3}=-0.460. \quad  \label{ps}
\end{eqnarray}
The wave functions of the left-handed quark zero modes and the first
excited state of $Q_1$ are shown in fig.~\ref{f_1}. 
We fix the relative positions of $Q_1$ and $Q_2$ by fitting $|V_{us}|$,
while $|V_{cb}|$ determines the relative positions of $Q_2$ and $Q_3$.
The Jarlskog invariant $J={\rm Im}(V_{cs}V_{us}^*V_{ud}V_{cd}^*)$ \cite{J} 
and $|V_{ub}|$ are then fixed as well. Note that the the CKM mixings
are determined by the locations of the left-handed quarks. We then use
the locations of the right-handed quarks to fit the quark masses. Taking
the quark locations (\ref{qlocations}), we find for the averages
\begin{eqnarray}
\begin{array}{lll} 
m_u=2.0~{\rm MeV},~~ & m_c=506~{\rm MeV},~~&  m_t=144~{\rm GeV},\\[.1cm]
m_d=4.0~{\rm MeV},~~ & m_s=58~{\rm MeV},~~& m_b=2.2~{\rm GeV},\\[.1cm]
|V_{us}|=0.222, & |V_{cb}|=0.040, &  |V_{ub}|=0.0088, \\[.1cm]
J=3.1\times10^{-5}.& &
\end{array}
\end{eqnarray} 
While the prediction of $J$ agrees well with the experimental value (\ref{J}),
$|V_{ub}|$ is found to be about two times too large (\ref{CKM}).
As discussed above this discrepancy is a consequence of the mass pattern
(\ref{product}). 

The quark locations of eq.~(\ref{qlocations}) are not a unique solution. Moving
all the left-handed quarks in such a way that the related factors $a_i$ in 
(\ref{product}) are changed by a common factor, can be compensated
by an appropriate change in the locations of the right-handed quarks. 
So in fact there is a one parameter family of solutions, labeled, for instance,
by $\delta c_{Q1}$. However, if some of
the light quarks are localized towards the TeV-brane ($c<1/2$), large deviations
from electroweak precision variables drive the KK scale far above
10 TeV \cite{HLS}. In section 5 we will see that in this case flavor
violation is greatly enhanced. Since $c_{U2}$ is already close to
1/2, we cannot move the right-handed quarks much closer to the TeV-brane.
This would also require to localize the right-handed top quark extremely
close towards the TeV-brane by $c_{U3}\lsim-2$.
On the other hand, the $Q_i$ cannot be localized much closer
towards the TeV-brane without getting too large modifications of
the left-handed bottom couplings. Thus we end up with the quite 
constrained range of $-0.01\lsim \delta c_{Q1}\lsim 0.02$. The quarks
can be localized closer towards the Planck-brane if the 5D Yukawa
couplings are increased by a common factor. Increasing them by
a factor of 10, for instance, corresponds to  $\delta c_{Q1}=0.037$.
However, very large Yukawa couplings $l\gg1$ introduce a new and
unexplained hierarchy in the model. Taking Yukawa couplings $l\ll1$
would mean to move the quarks closer towards the TeV-brane which
worsens the electroweak fit. The quark locations also depend 
on $k/M_{\rm Pl}$ which we take to be one.

The experimental errors (\ref{qmasses}) and (\ref{CKM}) translate
into uncertainties in the fermion locations. For instance, from $|V_{cb}|$
we can fix the position of $Q_3$ relative to $Q_2$ only
up to an error of $-0.02\lsim \Delta c\lsim 0.03$. The position 
of  $Q_1$ relative to $Q_2$ is determined quite accurately up to about
$-6\times10^{-4}\lsim \Delta c\lsim 5\times10^{-4}$ by $|V_{us}|$. For the
right-handed quarks the errors induced by the quark masses 
range from $\Delta c\approx\pm 8\times10^{-4}$
for $D_3$ to $\Delta c\approx\pm 0.03$ for $U_3$.
More precisely, these are errors on the relative positions of left- and 
right-handed quarks, for instance, $c_{Q3}-c_{U3}$ in the case of the top. 
Note that positions $c<1/2$ have larger errors since in this range 
the fermion mass does not depend exponentially but as $\sqrt{1/2-c}$ 
on the fermion location.

\begin{figure}[t] 
\begin{picture}(100,160)
\put(0,0){\epsfxsize7cm \epsffile{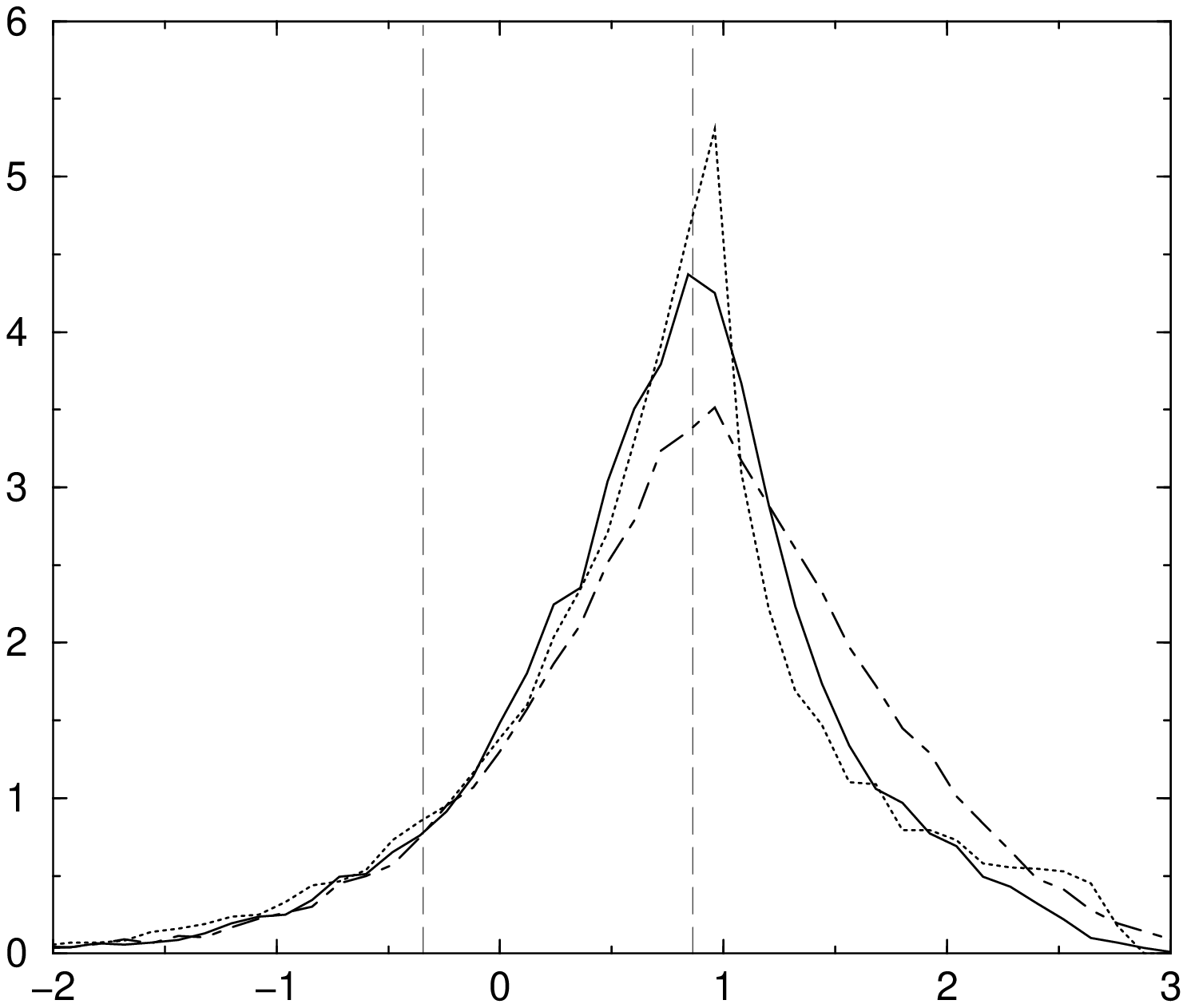}}
\put(220,0){\epsfxsize7cm \epsffile{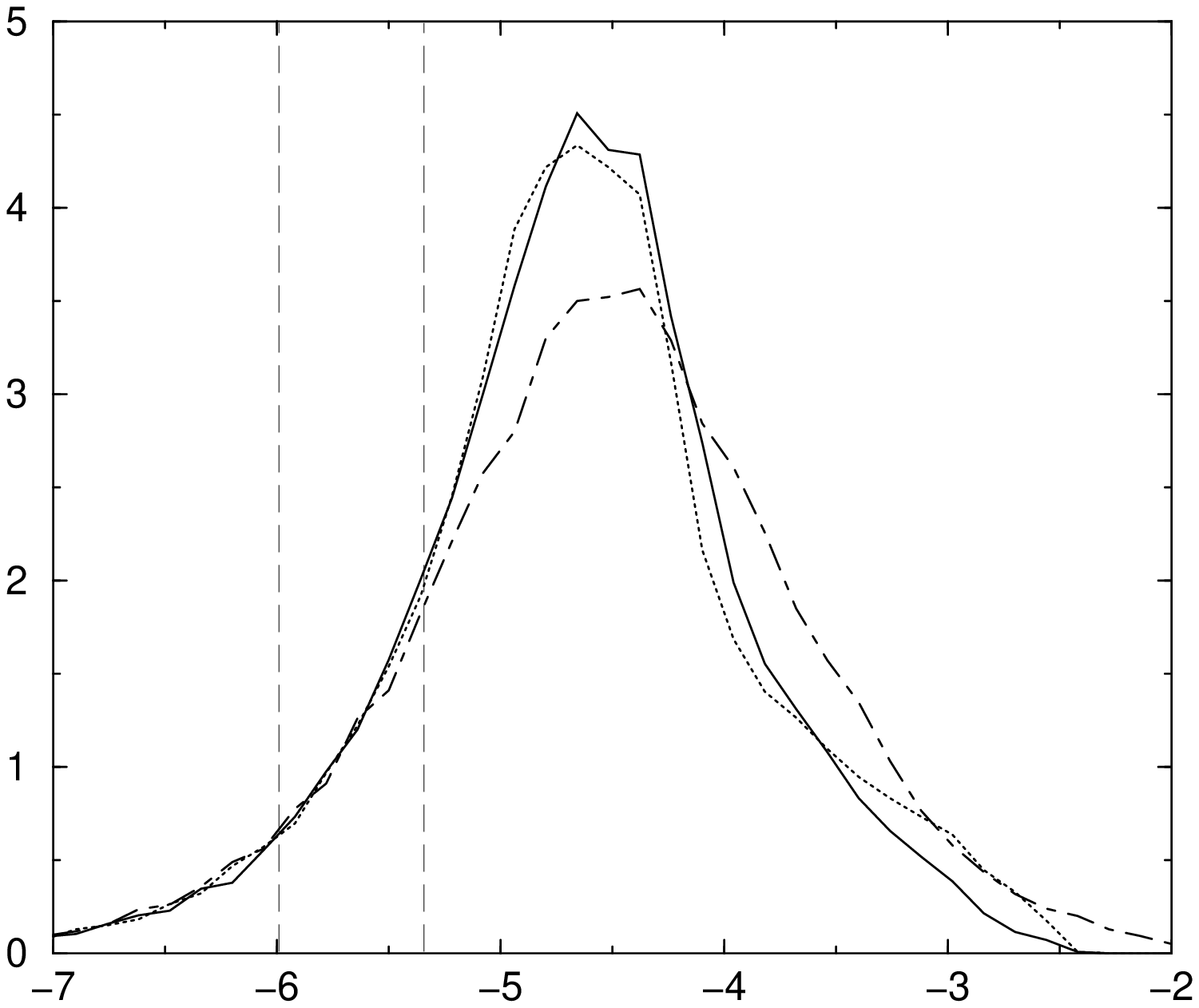}}
\put(205,115){\#{\large $\uparrow$}} 
\put(130,-10){$\ln(m_u[{\rm MeV}])\rightarrow$}
\put(350,-10){$\ln(|V_{ub}|)\rightarrow$}
\put(20,147){{(a)}} 
\put(390,147){{(b)}} 
\end{picture} 
\caption{Logarithmic distributions of $m_u$ (a) and $|V_{ub}|$ (b) for Yukawa
coupling distributions $2/3<|l_{ij}|<4/3$ (solid lines), $0<|l_{ij}|<2$ (dashed dotted lines) 
and $|l_{ij}|=1$ (solid lines). The vertical dashed lines indicate the experimental 
uncertainties.
}
\label{f_2}
\end{figure}

The statistical approach provides us with distributions of fermion
properties rather than with their precise values. This adds further
uncertainties to the fermion locations.  In fig.~\ref{f_2}a we
present the distribution of $\ln(m_u)$ which shows a half width of about 1.1.
Requiring that, with a different location, the average value of $\ln(m_u)$ 
still lies within this half width range translates into an uncertainty of
$-0.017\lsim \Delta c_{U1}\lsim 0.019$. In table \ref{t_error} we have
summarized the statistical and experimental uncertainties in the quark 
locations.  Fig.~\ref{f_2}b shows the distribution of $\ln(|V_{ub}|)$ which
overlaps considerably with the experimentally favored range.
Despite the factor of two deviation in the average $\langle|V_{ub}|\rangle$,
there is an acceptably good fit to the data once the statistical
variations are taken into account.

Our results depend only weakly on the range taken for the 
moduli of the 5D couplings $l_{ij}$. In fig.~\ref{f_2} we 
present the distributions of $m_u$ and
$|V_{ub}|$ additionally for the case $0<|l_{ij}|<2$ and for pure phase Yukawa 
couplings $|l_{ij}|=1$.  In the context of flat extra dimensions
pure phase Yukawa couplings have recently been investigated in
ref.~\cite{HS01}. We find that a wider range of $|l_{ij}|$ leads only
to a slight broadening in the distributions of fermion properties. The
induced shift in the average values is always much smaller than
the width of the distributions. For instance, we find an average
$m_u=2.3~(2.0)$ MeV in the case $0<|l_{ij}|<2~(|l_{ij}|=1)$. This
behavior stresses that the fermion mass pattern is indeed determined
by the fermion locations, not by accidental properties of the
5D Yukawa matrices.
Note that the robustness of our results concerning different
ranges of  $|l_{ij}|$ is due to the complex phases in the 5D Yukawa 
couplings. For real values of $l_{ij}$ there is a high probability of cancellations 
in the mass matrices. This leads to multiple peaks in the distributions 
of fermion masses and mixings for a narrow range of $|l_{ij}|$.

\begin{table}[b] \centering
{\small
\begin{tabular}{|c||c|c|c|c|c|c|c|c|c|} \hline
& ${Q_1} $ & ${Q_2} $ & ${Q_3}$ & $D_1$ & $D_2$ & $D_3$& $U_1$ & $U_2$ & $U_3$  \\ \hline 
$c$&0.643 &0.583 &0.317 &0.643 & 0.601 & 0.601 & 0.671& 0.528&--0.460 \\ \hline
exp.& +0.0004 & +0.002 & +0.03 &+0.017& +0.012 &+0.0008 & +0.017 & +0.004& +0.03\\ 
& --0.0006& --0.003& --0.02& --0.008 &--0.006& --0.0008  & --0.002& --0.002& --0.03\\ \hline
stat.& +0.022& +0.029 & +0.13&+0.016&+0.006& +0.004  & +0.019& +0.006 &+0.50 \\
&--0.027 & --0.024 & --0.21& --0.023 &--0.017&--0.006  & --0.017& --0.023&--0.76\\ \hline
\end{tabular}
} 
\caption{Experimental and statistical errors on the fermion locations
of eq.~(\ref{qlocations}).
 }
\label{t_error} 
\end{table}

Let us close this section with a brief discussion of the lepton sector.
To determine the lepton locations we have to take into account neutrino
masses and mixings. Neutrino masses can arise from different 
sources in the context of warped geometry. In one scenario sub-eV
Dirac masses are generated by a coupling to right-handed neutrinos
in the bulk \cite{GN,HS3}. The necessary tiny Yukawa couplings
can naturally be generated if the right-handed neutrinos are localized
closely towards the Planck-brane by taking $1.2\lsim c\lsim 1.5$.  
Neutrino and charged fermion masses are then treated on the same 
footing. Alternatively, the dimension five interaction $(1/Q)HHLL$
can induce small Majorana neutrino masses \cite{HS4}. The suppression scale
$Q$ of the non-renormalizable interaction depends on the position
of the left-handed leptons $L$ in the extra dimension. This mechanism
is minimal in the sense that it does not require to extend the standard
model particle content. 

As in the case of quarks, the locations of the left-handed states determine 
the observed fermion mixings.  Large neutrino mixings require the
neutrinos and thus the SU(2) lepton doublets to have similar positions $c_{Li}$
\begin{eqnarray} \label{llocations1}
c_{L1}=0.565, \quad &c_{L2}=0.565, \quad  &c_{L3}=0.565, \quad  \nonumber \\
c_{E1}=0.764, \quad &c_{E2}=0.609, \quad &c_{E3}=0.509. \label{psl}
\end{eqnarray}
The right-handed positions $c_{Ei}$ we fixed by requiring that
with random Yukawa couplings the average charged lepton masses
fit their observed values. 

To suppress the matrix element $U_{e3}$  in the neutrino mixing matrix 
it is favorable to separate the
electron doublet somewhat from the muon and tau doublets \cite{HS4}
 \begin{eqnarray}  \label{llocations2}
c_{L1}=0.631, \quad &c_{L2}=0.565, \quad  &c_{L3}=0.565, \quad  \nonumber \\
c_{E1}=0.725, \quad &c_{E2}=0.594, \quad &c_{E3}=0.497. 
\end{eqnarray}
Like in the case of quarks there is a one parameter family of degenerate locations 
if the left- and right-handed states are shifted against each other.  The widths
of the distributions allow to fix the locations only up to $\Delta c\approx \pm 0.01$.
Taking the 5D Yukawa couplings larger than one, we could shift
the leptons closer towards the Planck-brane, as was done in ref. \cite{HS4}.

\section{Mixings of KK states and the unitarity of the CKM matrix}
So far we have only dealt with the zero modes of quarks and leptons. But
the Yukawa interaction (\ref{3.1}) also induces mixings between the zero modes
and the vector-like excited states. In the following we show that this effect
hardly modifies the conclusions we have reached above.

From the KK reduction of an SU(2) doublet quark $Q_L$
we obtain a left-handed zero mode $q_L^{(0)}$, corresponding to
the SM quark,  and an infinite
tower of left- and right-handed KK excited states $q_L^{(n)}$ and 
 $q_R^{(n)}$, where we omit flavor indices. The SU(2) singlet up
quarks decompose into the right-handed zero mode 
$u_R^{c(0)}$ and the
KK excited states $u_L^{c(n)}$ and $u_R^{c(n)}$. After electroweak
symmetry breaking the up quark mass matrix takes the form
\begin{equation} \label{KK_mass}
M_{U}= (\bar u_L^{(0)},\bar u_L^{(1)},\bar u_L^{c(1)},\dots) \left(\begin{array}{cccc} 
m^{(0,0)} & 0 &m^{(0,1)} & \cdots \\[.1cm] 
m^{(1,0)} & m_{Q,1} & m^{(1,1)}& \cdots \\[.1cm]  
0 & 0 & m_{U,1} & \cdots \\
\vdots & \vdots & \vdots & \ddots
\end{array}\right)\left(\begin{array}{c}u_R^{c(0)} \\[.1cm]   u_R^{(1)} \\[.1cm]   
u_R^{c(1)} \\ \vdots \end{array}\right) 
\end{equation} 
where we again suppress flavor indices, i.e.~every entry represents
a $3\times 3$ matrix in flavor space. The masses $m^{(m,n)}$ arise from
electroweak symmetry breaking. They
are obtained by inserting the relevant wave functions
into eq.~(\ref{3.2}). The $m_{Q,m}$ and $m_{U,m}$ denote the KK masses
of the excited quark states. The zeros in (\ref{KK_mass}) follow 
from the $Z_2$ orbifold properties of the wave functions.
Analogous mass matrices arise for the down quarks and charged
leptons. The wave functions of the the KK fermions are localized
at the TeV-brane (see fig.~\ref{f_1}). Their overlap with the Higgs is large and
almost independent of the location of the zero mode. We obtain
\begin{eqnarray}
m^{(0,n)}&=&\frac{lv_0}{\pi kR}f_{0L}(\pi R)f_{nR}(\pi R)\nonumber \\[.2cm]
m^{(m,n)}&=&\frac{lv_0}{\pi kR}f_{mL}(\pi R)f_{nR}(\pi R)\approx (-1)^{m+n}
\cdot l\cdot 349{\rm ~GeV},~~m,n\geq1.
\end{eqnarray}
These masses are large compared to the corresponding zero mode
masses.

The mass matrix (\ref{KK_mass}) induces a mixing
between SU(2) singlet and doublet states. This effect diminishes
the weak charge of the left-handed quarks \cite{dAS}. In the context of
neutrinos this  behavior was discussed in refs.~\cite{GN,HS3}. 
Ignoring flavor mixing for a moment, the singlet admixture in
the left-handed zero mode is
\begin{equation} 
\sum_{n=1}^{\infty}\left(\frac{m^{(0,n)}}{m_{U,n}}\right)^2 
\approx \left(\frac{m^{(0,1)}}{m_{U,1}}\right)^2 \sum_{n=1}^{\infty}\frac{1}{n^2}
=\left(\frac{m^{(0,1)}}{m_{U,1}}\right)^2 \frac{\pi^2}{6}.
\nonumber
\end{equation} 
In the second step we approximated the KK spectrum by $m_n\approx nm_1$.
Taking the quark positions of eq.~(\ref{qlocations}) we obtain
for the top quark a singlet admixture of $1.5\times 10^{-4}$.
For the other quark flavors the admixture is smaller than $10^{-6}$.
The mixing between the SU(2) doublet zero mode and its KK excitations is
even more suppressed and on the order of $(m^{(0,1)}m^{(1,1)})^2/m_{Q,1}^4$. 
Our neglect of the KK states in the discussion
of quark and lepton masses in the previous section is therefore
well justified. 

\begin{figure}[t] 
\begin{picture}(100,160)
\put(0,5){\epsfxsize7cm \epsffile{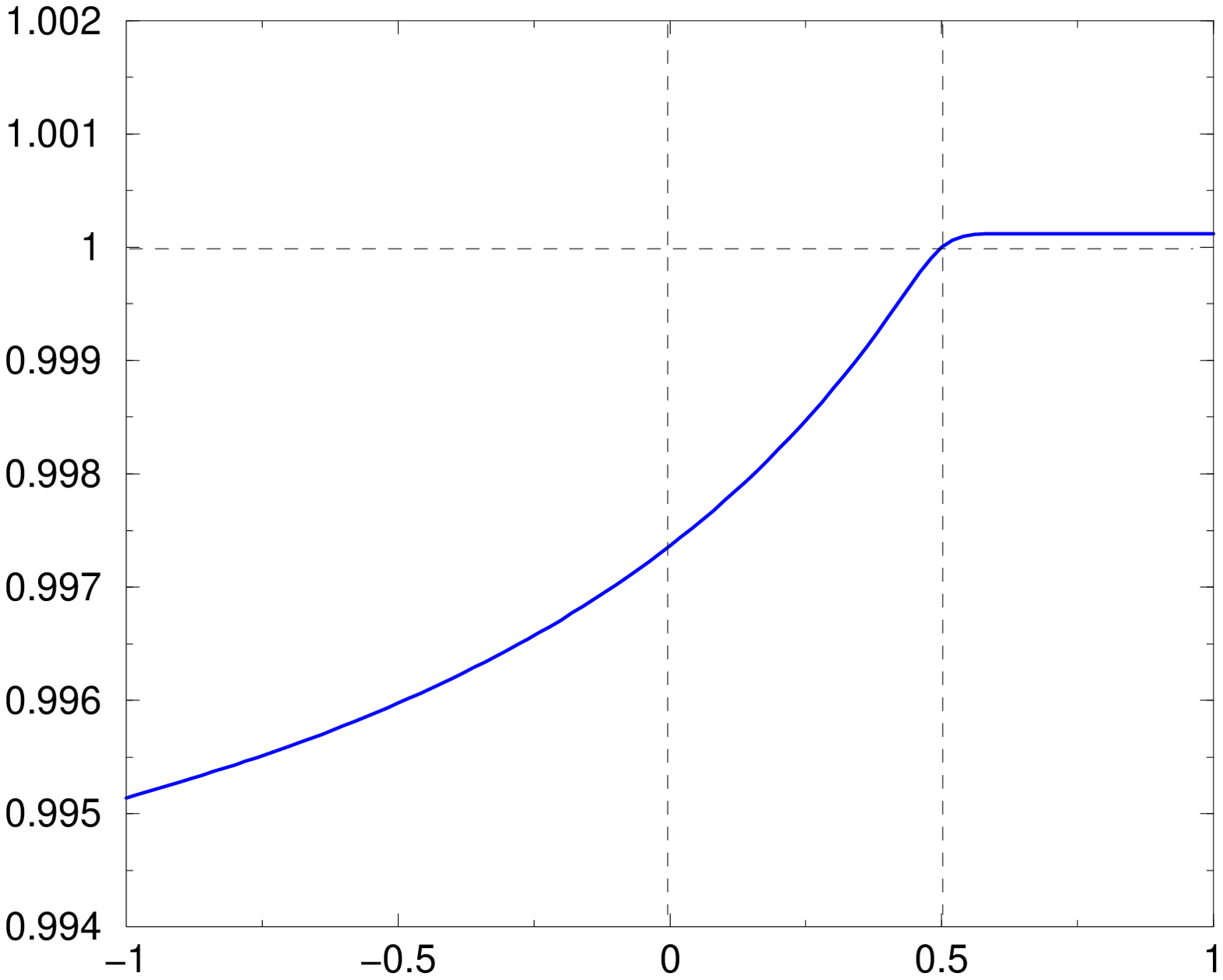}}
\put(220,5){\epsfxsize6.8cm \epsffile{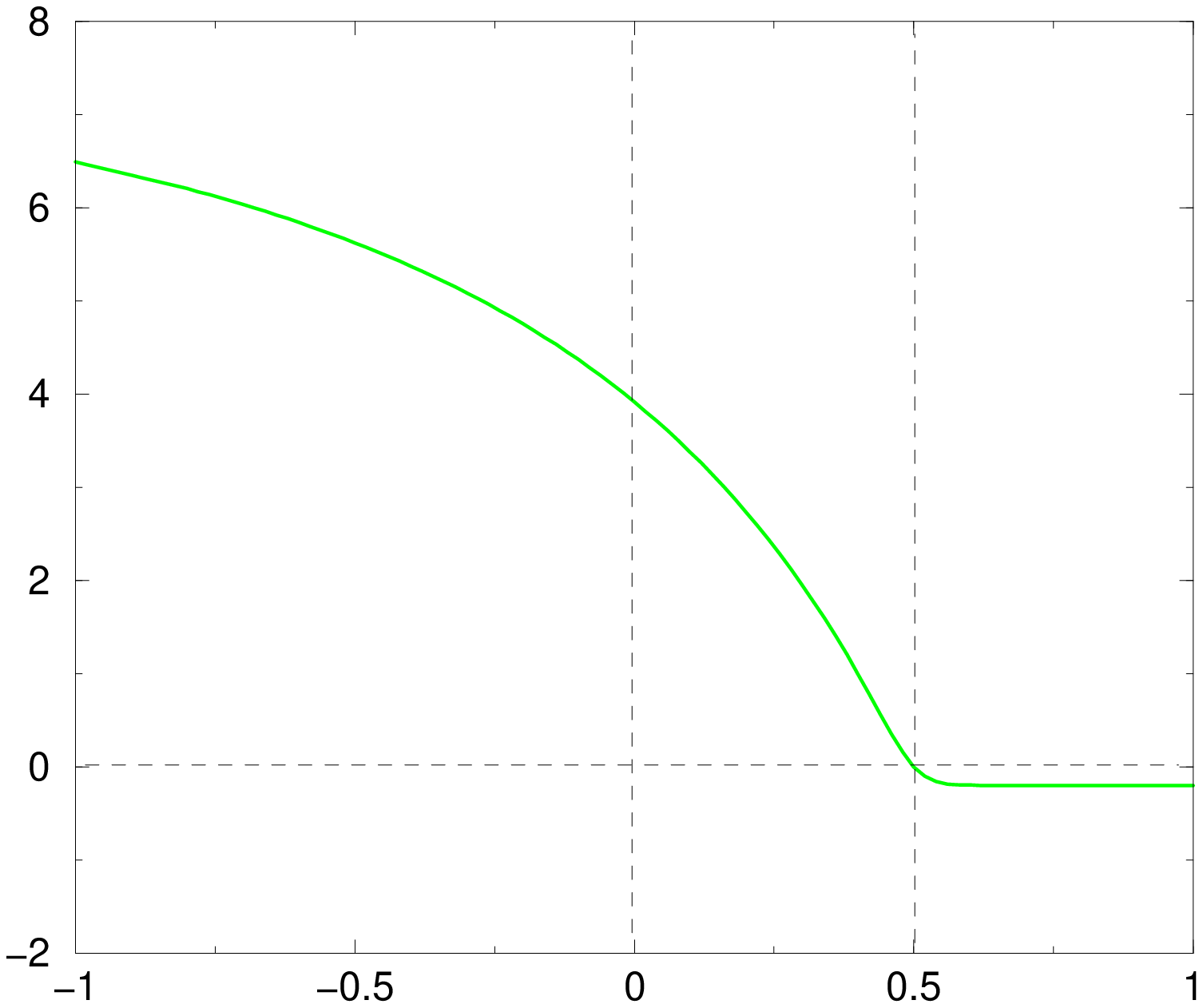}}
\put(50,60){{$\frac{g_0}{g_{SM}}$}} 
\put(260,110){{$\frac{g_1}{g_{SM}}$}} 
\put(130,-7){$c\rightarrow$}
\put(350,-7){$c\rightarrow$}
\put(25,147){{(a)}} 
\put(390,147){{(b)}} 
\end{picture} 
\caption{Gauge couplings of the Z boson (a) and its first KK
state relative to the SM value as a function of the fermion
location.
}
\label{f_3}
\end{figure}

The gauge couplings of fermions (\ref{gc}) are modified by KK mixing
in the weak gauge boson sector \cite{HS,HLS}. The W and Z bosons
couple somewhat weaker to fermions at the TeV-brane, while their
KK states couple stronger \cite{DHRP,GP}, as we show in fig.~\ref{f_3}. 
For $c>1/2$ the gauge coupling is almost independent of the precise 
location, and the coupling of the KK states is small, $g_1/g_{SM}\approx0.19$.

The charged current interaction of the KK fermions is described
by an infinitely dimensional version of the CKM matrix 
\begin{equation} \label{4.30}
{\cal V}={\cal U}_{L,U} {\cal G}{\cal U}_{L,D}^{\dagger} 
\end{equation}
where ${\cal U}_{L}$ and ${\cal U}_{R}$ diagonalize the full 
fermion mass matrices (\ref{KK_mass}). The flavor diagonal matrix
${\cal G}$ contains the gauge couplings of the various KK
states to the W boson. We  normalize ${\cal G}$ to the weak coupling 
of the muon as obtained from eq.~(\ref{gc}) with the muon location
of eq.~(\ref{llocations1}).
Since the weak gauge bosons are massive, they induce
couplings between different KK levels. These couplings are
suppressed by $M_Z/M_{KK}$ and numerically small. Note that
in general even the full CKM matrix $\cal V$ is not unitary. Its truncation to 
the zero mode sector, $V$, governs the low energy charged
current interaction. Three sources contribute to the non-unitarity of $V$:
the mixing of SU(2) doublet and singlet states, the truncation to the zero 
mode sector and the modification of the fermion gauge couplings.
The last contribution is dominant, except for the third generation
quarks. Defining  $\Delta V_{i}=\sum_{j=1}^3|V_{ij}|^2-1$, we find
with the quark locations of eq.~(\ref{qlocations}) 
$\Delta{V}_1\approx1\cdot10^{-5}$,
$\Delta{V}_2\approx2\cdot10^{-7}$ and
$\Delta{V}_3\approx-3\cdot10^{-3}$.
Again we have averaged over random sets of Yukawa couplings. 
The deviation from unitarity in the first generation is two orders
of magnitude below the current experimental sensitivity \cite{PDG}.
Also the expectation of $\Delta{V}_3$ is small compared to the
few percent precision in the weak charge of the top quark to be
reached at the LHC \cite{beneke}.  

Charged current interactions are also mediated by the KK states of
the W boson. These corrections are suppressed by
$(g_1^2/\delta g_0^2)(M_W^2/M_{KK}^2)\lsim 0.1$ and therefore 
subleading. 
The inequality becomes satisfied for the top quark which is localized
closest towards the TeV-brane. Only for fermions localized very closely
to the TeV-brane these corrections become important.

\section{Flavor violation}
The mechanisms we have discussed in the previous section in the
context of charged current sector modify the neutral current
interactions in a similar way. In analogy to eq.~(\ref{4.30}) we define 
the neutral current gauge couplings in the basis of mass eigenstates as
\begin{equation} \label{nc}
{\cal X}_{L,R}^{\psi (n)}={\cal U}_{L,R}^{\psi} {\cal G}_{L,R}^{\psi (n)}
{\cal U}_{L,R}^{\psi\dagger} 
\end{equation} 
where the unitary matrices ${\cal U}_{L,R}^{\psi}$ diagonalize the full
mass matrices, including the KK states of the fermion species 
$\psi=q_L, u_R,l_L,$ etc. In ref.~\cite{H02} we had not yet included
KK mixing in the fermion sector. 
With ${\cal G}_{L,R}^{\psi n}$ we denote
the coupling to the $n$th KK state of the Z boson as obtained from
eq.~(\ref{gc}). If the fermion families are localized at different positions
in the extra dimension, the gauge couplings  ${\cal G}_{L,R}^{\psi n}$
are non-universal in flavor space. These deviations in the fermion
gauge couplings are too small to directly show up in flavor conserving
processes at present accelerator experiments if the KK scale is about
10 TeV \cite{HLS}. However, the transformation to the mass eigenstates (\ref{nc})
induces flavor violating couplings which lead to interesting new
phenomena. These flavor violating interactions are an immediate
consequence of our approach to the problem of fermion masses. The phenomenology
is similar to what happens in models with family non-universal  Z' bosons, 
so we can adopt the formalism described, for instance, in ref.~\cite{LP}
with the complication of the KK mixing of fermions. In the gauge boson
sector we work with mass eigenstates from the very beginning. There
is no mixing of the Z boson and its KK states, instead the 
zero mode has flavor violating couplings as well.    

Flavor violation as induced by the couplings (\ref{nc}) is driven by the
non-universality $\delta g_{ij}$ in the couplings of different flavor states 
which are mixed by an angle $\theta _{ij}$
\begin{equation} 
{\cal X}_{ij}\approx \delta g_{ij} \sin\theta_{ij}.
\end{equation} 
Very different fermion locations on the one hand increase 
non-universality, on the other hand they lead to small mixing
angles. Non-universality is larger for fermions localized towards
the TeV-brane, $c<1/2$, as is shown in fig.~\ref{f_3}, and
for the Z zero mode depends on the KK scale as $1/M_{KK}^2$.
Note that, different from the standard model, the 
right-handed mixings become physically relevant.  
The left-handed mixings of quarks and leptons are naturally 
comparable to the observed CKM and neutrino mixing angles, 
respectively. This requirement fixed the left-handed fermion locations 
in section 3. The locations of right-handed fermions we have used to fit
the fermion masses.  For the locations (\ref{qlocations}) and 
(\ref{llocations2})  the right-handed mixings are then
predicted as
\begin{equation}\label{rightmxings}
\begin{array}{lll} 
\langle |{\cal U}_{R,1,2}^u|\rangle=0.017, &\langle |{\cal U}_{R,2,3}^u|\rangle=0.073,
&\langle |{\cal U}_{R,1,3}^u|\rangle=0.0012, \\[.2cm]
\langle |{\cal U}_{R,1,2}^d|\rangle=0.23, &\langle |{\cal U}_{R,2,3}^d|\rangle=0.60,
&\langle |{\cal U}_{R,1,3}^d|\rangle=0.23, \\[.2cm]
\langle |{\cal U}_{R,1,2}^e|\rangle=0.016, &\langle |{\cal U}_{R,2,3}^e|\rangle=0.057,
&\langle |{\cal U}_{R,1,3}^e|\rangle=0.0016.
\end{array}
\end{equation}
Again we have averaged over random sets of Yukawa couplings.
The right-handed down quarks mix strongly with each other, while
the mixings of up quarks and leptons are found to be small. For the 
alternative other set of lepton locations (\ref{llocations1}) the right-handed
mixings are similarly small.

In the following we are using a KK scale of 10 TeV which might be 
somewhat lowered by introducing brane-kinetic 
terms \cite{branekin}. Note that the rates of flavor violating processes 
which we present below cannot be translated to that case by simple 
scaling with the appropriate powers of $M_{KK}$. Brane-kinetic terms
modify the KK wave functions and thus possible non-universalities even
if the KK scale is not changed. 

\subsection{Lepton flavor violation}
Let us start the discussion of flavor violation with the lepton sector.
In addition to the lepton locations (A) (\ref{llocations1}) and (B) (\ref{llocations2})
we are considering two modified settings
\begin{eqnarray} \label{llocations3}
{\rm (A'):}~~c_{L1}=0.585, \quad &c_{L2}=0.565, \quad  &c_{L3}=0.545, \quad  
\nonumber \\
{\rm (B'):}~~c_{L1}=0.520, \quad &c_{L2}=0.200, \quad  &c_{L3}=0.150 \quad  
\end{eqnarray}
to explore the dependence on the fermion locations.
The right-handed positions are taken as in (A) and (B), respectively.
In the set (A') we have shifted the left-handed leptons by a small
amount to induce some non-universality in the left-handed sector. 
The changes in the lepton 
masses and mixings remain within the widths of the statistical 
distributions. In (B') we have reduced the Yukawa couplings by 
a common factor of twenty, which is compensated by moving 
the left-handed leptons closer towards the TeV-brane where
non-universality is especially large. We also separated somewhat
the left-handed muon and tau states. Using these locations we
find the following lepton flavor violation Z couplings
\begin{equation} \label{lfvc}
\begin{array}{ccccc} 
&{\rm (A)} & {\rm (A')} & {\rm (B)} & {\rm (B')} \\[.2cm]
\langle |{\cal X}_{L,1,2}^{e(0)}|\rangle: &
2.3\times10^{-7} & 6.5\times10^{-7} & 3.7\times10^{-7} & 8.4\times10^{-5} 
\\[.2cm]
\langle |{\cal X}_{L,2,3}^{e(0)}|\rangle: &
5.1\times10^{-7} & 1.2\times10^{-6} & 3.5\times10^{-7} & 3.2\times10^{-5} 
\\[.2cm]
\langle |{\cal X}_{L,1,3}^{e(0)}|\rangle: &
2.3\times10^{-7} & 8.5\times10^{-7} & 2.2\times10^{-7} & 4.9\times10^{-5} 
\\[.2cm]
\langle |{\cal X}_{R,1,2}^{e(0)}|\rangle: &
7.1\times10^{-9} & 9.7\times10^{-9} & 1.0\times10^{-8} & 7.7\times10^{-9} 
\\[.2cm]
\langle |{\cal X}_{R,2,3}^{e(0)}|\rangle: &
7.6\times10^{-7} & 1.1\times10^{-6} & 2.3\times10^{-6} & 1.7\times10^{-6} 
\\[.2cm]
\langle |{\cal X}_{R,1,3}^{e(0)}|\rangle: &
7.3\times10^{-9} & 8.5\times10^{-9} & 6.5\times10^{-8} & 4.6\times10^{-8}
\end{array}
\end{equation}
where we have included the zero modes and the first KK states of the leptons.
Note that in the case (B') there is a large increase in the left-handed 
couplings compared to the cases with $c\gsim 1/2$. For (A) the flavor
violating left-handed couplings completely stem from fermion KK mixing 
since there is no non-universality among the left-handed states. In the other
cases fermion KK mixing is a subleading effect.

The couplings (\ref{lfvc}) induce lepton flavor violating decays of the Z boson
with a branching ratio ${\rm Br}(Z\rightarrow l_i\bar l_j)\approx0.29\cdot (|{\cal X}_{L,i,j}^{e(0)}|^2+|{\cal X}_{R,i,j}^{e(0)}|^2)$. The largest rate
occurs in scenario (B') with $(Z\rightarrow e\bar \mu)\approx 2\times 10^{-9}$
which is still three orders of magnitude below the experimental bound
of $ 1.7\times 10^{-6}$ \cite{PDG}. The other decay modes are even stronger
suppressed. Thus there are no constraints on the model from lepton flavor violating
Z decays.

The off-diagonal elements of ${\cal X}$ also lead to flavor violating decays 
of charged leptons. Tree-level exchange of a Z boson and its 
KK states induces the processes $l_i\rightarrow 3l_j$ at a rate
\begin{equation} \label{mu3e}
\begin{array}{rcccc} 
&{\rm (A)} & {\rm (A')} & {\rm (B)} & {\rm (B')} \\[.2cm]
{\rm Br}(\mu\rightarrow eee): &
5.2\times10^{-14} & 4.7\times10^{-13} & 1.0\times10^{-13} & 5.3\times10^{-9} 
\\[.2cm]
{\rm Br}(\tau\rightarrow \mu\mu\mu): &
1.1\times10^{-13} & 2.1\times10^{-12} & 7.3\times10^{-13} & 1.7\times10^{-10} 
\\[.2cm]
{\rm Br}(\tau\rightarrow eee): &
7.5\times10^{-15} & 3.2\times10^{-13} & 7.8\times10^{-15} & 3.9\times10^{-10}. 
\\[.2cm]
\end{array}
\end{equation}
Again we have averaged over random lepton Yukawa couplings 
in the four different scenarios. In the result (\ref{mu3e}) we have 
only included the contribution of the Z boson exchange. The amplitude 
contributed by its first KK state is already suppressed by a factor of about twenty. 
We have included the first KK level of lepton states which is important  
in scenario (A) where the left-handed lepton positions are degenerate.
The branching ratios depend on the KK scale as $1/M_{KK}^4$.
The experimental bound ${\rm Br}(\mu\rightarrow eee)<1.0\times10^{-12}$ 
\cite{mu3e} is satisfied for leptons localized towards
the Planck-brane and the rates can come close to the experimental
sensitivity. It would be very interesting if a slow muon facility could 
test this process at the precision of $10^{-16}$ \cite{SMF}.
The scenario (B'), where the leptons are moved
towards the TeV-brane, leads to strong flavor violation. The rate could
only be brought down to the experimental bound if the KK scale is increased 
by an order of magnitude to about 100 TeV. The expected rates for tau decays 
are in all cases well below the latest Belle results 
${\rm Br}(\tau\rightarrow \mu\mu\mu)<3.8\times10^{-7}$ and 
${\rm Br}(\tau\rightarrow eee)<2.7\times10^{-7}$ \cite{tau3e}.

Severe experimental bounds have been put on $\mu e$-conversion
in muonic atoms. The best exclusion limit 
${\rm Br}(\mu N\rightarrow e N)\equiv \Gamma(\mu N\rightarrow e N)/\Gamma(\mu N\rightarrow \nu_{\mu} N')<6.1\times10^{-13}$ comes from the Sindrum-II collaboration
\cite{SindrumII}. We expect a branching ratio of
\begin{equation} \label{mue}
\begin{array}{ccccc} 
&{\rm (A)} & {\rm (A')} & {\rm (B)} & {\rm (B')} \\[.2cm]
{\rm Br}(\mu N\rightarrow e N): &
5.0\times10^{-16} & 5.2\times10^{-15} & 1.0\times10^{-15} & 5.8\times10^{-11}.
\end{array}
\end{equation}
While the scenario (B') is again excluded for a KK scale of 10 TeV,
leptons close towards the Planck-brane easily avoid the experimental
limit. The forthcoming MECO experiment (E940 at BNL) plans to
increase the sensitivity to about $5\times 10^{-17}$ \cite{bachmann}. 
It would completely cover the predicted interaction rates, even for 
leptons close to the Planck-brane.

At the one-loop level radiative decays that violate lepton flavor are 
induced at a rate
\begin{equation} \label{muegamma}
\begin{array}{ccccc} 
&{\rm (A)} & {\rm (A')} & {\rm (B)} & {\rm (B')} \\[.2cm]
{\rm Br}(\mu\rightarrow e\gamma): &
2.1\times10^{-16} & 1.9\times10^{-15} & 4.1\times10^{-16} & 2.1\times10^{-11} 
\\[.2cm]
{\rm Br}(\tau\rightarrow \mu\gamma): &
6.7\times10^{-16} & 9.5\times10^{-15} & 4.2\times10^{-15} & 6.6\times10^{-13} 
\\[.2cm]
{\rm Br}(\tau\rightarrow e\gamma): &
2.8\times10^{-17} & 1.3\times10^{-15} & 3.3\times10^{-17} & 1.5\times10^{-12}. 
\\[.2cm]
\end{array}
\end{equation}
Except for the scenario (B') the expected branching ratios are far below
the experimental bounds ${\rm Br}(\mu\rightarrow e\gamma)<1.2\times 10^{-11}$,
${\rm Br}(\tau\rightarrow \mu\gamma)<1.1\times10^{-6}$ and
${\rm Br}(\tau\rightarrow e\gamma)<2.7\times10^{-6}$. The MEG collaboration
at PSI plans to increase the sensitivity with respect to $\mu\rightarrow e\gamma$ to about
$10^{-13}-10^{-14}$ within the next years \cite{MEG}. This will not be sufficient
to test the model for leptons localized towards the Planck-brane.
Note that our model leads to predictions quite different from supersymmetric models, 
where radiative decays occur at much larger rates than the decays of 
eqs.~(\ref{mu3e}) and (\ref{mue}) \cite{CIHN}. In the warped SM 
radiative decays occur only
at the loop level, while the latter are tree-level processes.
In the scenario of Dirac neutrino masses radiative decays  
are also mediated by the KK states of the sterile neutrinos. 
If the SM neutrinos are confined to the TeV-brane,
a too large branching ratio for $\mu\rightarrow e\gamma$ 
pushes the KK scale up to 25 TeV and thus imposes a
stringent constraint on the model \cite{K00}.  
However, the rate for $\mu\rightarrow e\gamma$ is 
sensitive to the mixing between light and heavy neutrino states.
With bulk neutrinos the mixing with heavy states is considerably
reduced. In ref.~\cite{HS3} a branching ratio 
Br$(\mu\rightarrow e\gamma)\approx 10^{-15}$ was found 
for leptons localized towards the Planck-brane ($c>1/2$).
While this value is still well below the experimental sensitivity,
it is larger than the contribution from gauge
boson exchange (\ref{muegamma}).
Other lepton flavor violating processes, such as muonium-antimuonium
oscillations \cite{willmann98}, do not lead to new constraints.

\subsection{Meson mass splittings and CP violation}
The flavor violating gauge couplings (\ref{nc}) also contribute to the
mass splittings in neutral pseudo-scalar meson systems. The mass
splitting between the flavor eigenstates of a meson, $P^0$ and $\bar P^0$,
is given by 
\begin{equation} \label{delta_m}
\Delta m_P=\frac{{\rm Re}\langle P^0|-{\cal L}_{\rm FCNC}|\bar P^0\rangle}{m_K}
\end{equation}
where ${\cal L}_{\rm FCNC}$ contains the flavor violating couplings (\ref{nc}).
Phases in the flavor violation couplings are constrained by the indirect 
CP-violation in the kaon system 
\begin{equation}\label{epsilon_K}
\epsilon_K=\frac{{\rm Im}\langle P^0|-{\cal L}_{\rm FCNC}|\bar P^0\rangle}
{2\sqrt{2}m_K\Delta m_H}.
\end{equation}
These expressions can be evaluated using the vacuum insertion 
approximation (see for instance \cite{VIA}).  The largest contributions 
to (\ref{delta_m}) and (\ref{epsilon_K}) come from the exchange of KK
gluons \cite{DPQ99}. Excited gluons have flavor non-universal couplings
to fermions like the KK states of the weak gauge bosons and induce
flavor violating couplings analogous to those of eq.~(\ref{nc}). The gluon zero 
modes couple universally and therefore do not mediate flavor changing
interactions.  

For the quark locations of eq.~(\ref{qlocations}) we obtain the following 
contributions of first level of KK gluons and the Z boson zero mode
\begin{equation} \label{meson}
\begin{array}{rccc} 
&{\rm gluon} & {\rm Z} & {\rm exp.}  \\[.2cm]
\Delta m_K: & 1.5\times10^{-14} & 1.2\times10^{-17} & 3.5\times10^{-12} \\[.2cm]
\Delta m_B: & 5.1\times10^{-11} & 3.0\times10^{-14} & 3.2\times10^{-10} \\[.2cm]
\Delta m_D: & 3.8\times10^{-13} & 5.2\times10^{-15} & 4.6\times10^{-11} \\[.2cm]
\epsilon_K: & 1.1\times10^{-3} & 1.1\times10^{-6} & 2.3\times10^{-3}.
\end{array}
\end{equation}
The meson mass splittings are given in units of MeV. In eq.~(\ref{meson})
we give results averaged over random sets of Yukawa couplings
$2/3<|l_{ij}|<4/3$.
As in the case of flat extra dimensions \cite{DPQ99} the dominating contributions
come from the exchange of KK gluons. KK states of the Z boson give only
an about ten percent correction to the zero mode exchange.
In the computation of the
fermion mass eigenstates we have included the first level of KK states. Compared
to using only fermion zero modes the gluon contribution to $\epsilon_K$ 
is enhanced by a factor of two. The meson mass splittings are less sensitive
to whether or not KK fermions are included. The neglect of higher KK levels
amounts to an uncertainty of order unity in our results. 

The contributions to meson mass splittings we find are much smaller than
the experimental values \cite{PDG} listed in the last column of  eq.~(\ref{meson}).
The KK gluon contribution to $\epsilon_K$ comes rather close to the 
observed value. About one third of the random sets of Yukawa
couplings we tested gave $\epsilon_K>2.3\times10^{-3}$, which is still 
acceptable. 
In ref.~\cite{KKS02} KK contributions to $b\rightarrow s\gamma$ were 
considered. For the quark locations of eq.~(\ref{qlocations}) a bound
on the KK scale of about 5 TeV was found. From the the process
$K^+\rightarrow\pi^+\nu\bar\nu$ the limit on the Z coupling
$|{\cal X}_{L,1,2}^{d(0)}|<5.1\times10^{-6}$ was obtained in ref.~\cite{B02}.
We find a considerably lower value of 
$\langle|{\cal X}_{L,1,2}^{d(0)}|\rangle=2.2\times10^{-7}$.
Our approach to fermion masses and mixings is therefore 
nicely compatible with experimental constraints on flavor violation.

We stress that our results are very different from models where
the fermions mass pattern is explained by flavor-dependent fermion 
locations in one or more universal {\em flat} extra dimensions. There 
flavor violation, in particular kaon mixing,  leads to very 
restrictive bounds on the KK scale, 
$M_{KK}\gsim10^3$ TeV \cite{DPQ99}, disfavoring these models
as a solution to the the gauge hierarchy problem.  Recently it was shown 
that models constructed from intersecting D-branes suffer from
a similar problem \cite{AMS03}. 
The crucial difference lies in the wave functions of the KK gauge bosons.
With a warped extra dimension the gauge boson wave functions are 
almost constant away from the TeV-brane. Therefore the gauge couplings
of KK gauge bosons are nearly universal for fermions localized somewhat
towards the Planck-brane ($c>1/2)$ as we show in fig.~\ref{f_3}. Flavor
violation is thus automatically suppressed for the light fermion species 
which are expected to reside closely towards the Planck-brane to 
explain their small masses. 

\section{Non-renormalizable operators and rare processes}
In models where the weak scale is identified with the fundamental 
scale of gravity, the low cut-off scale dramatically amplifies the impact of
non-renormalizable operators in weak scale interactions. 
As a consequence, large rates for rare processes, such as flavor 
violation and proton decay, are a challenge for model building. 
With bulk fermions localized towards the Planck-brane the corresponding 
suppression scales can be significantly enhanced without relying
on ad-hoc symmetries \cite{GP,HS2}.
However, there are limits because the SM fermions need
to have sufficient overlap with the Higgs field at the TeV-brane
to acquire their observed masses, as discussed in section 3. 
We consider the following generic four-fermion operators 
which are relevant for flavor violation as well as for proton decay 
\begin{equation}
\int d^4x \int dy \sqrt{-g}\frac{1}{M_5^3}\bar \Psi_i\Psi_j\bar\Psi_k\Psi_l
\equiv \int d^4x \frac{1}{Q^2}\bar \Psi_i^{(0)}\Psi_j^{(0)}\bar\Psi_k^{(0)}\Psi_l^{(0)}.
\end{equation}
Integrating over the extra dimension, the effective 4D suppression 
scales $Q$ associated with these 
operators depend on where the relevant fermion states are localized 
in the extra dimension. 

Let us focus on some examples. 
The lepton flavor violating decay $\mu\rightarrow eee$
is induced by the operator $\mu eee$ at a rate $\Gamma\sim m_{\mu}^5/Q^4$.
The experimental constraint on the corresponding branching ratio translates into
 $Q>5\times 10^{5}$ GeV. For the lepton locations considered in the previous
section we obtain the suppression scales
\begin{equation} \label{nrmueee}
\begin{array}{ccccc} 
&{\rm (A)} & {\rm (A')} & {\rm (B)} & {\rm (B')} \\[.2cm]
Q(\mu eee)[{\rm GeV}]: &
4.8\times10^{6} & 1.1\times10^{7} & 7.9\times10^{7} & 2.4\times10^{5},
\end{array}
\end{equation}
where we have taken all the leptons to be left-handed. Except for the 
case (B') where the leptons are localized towards the TeV-brane, the
operator $\mu eee$ is safely suppressed. The same holds for similar lepton 
flavor violating operators such as $\tau \mu\mu\mu$ and $\tau eee$.
The operator $\mu e q_1q_1$ contributing to muon electron conversion is
constrained by $Q>1\times 10^{5}$ GeV while we are finding
\begin{equation} \label{nrmu}
\begin{array}{ccccc} 
&{\rm (A)} & {\rm (A')} & {\rm (B)} & {\rm (B')} \\[.2cm]
Q(\mu eq_1q_1)[{\rm GeV}]: &
4.4\times10^{7} & 3.5\times10^{7} & 1.1\times10^{8} & 6.6\times10^{6}.
\end{array}
\end{equation}
These suppressions scales even exceed $Q>1\times 10^{6}$ GeV,
a bound which could be set by the upcoming MECO experiment. Thus 
non-renormalizable operators are not expected to induce lepton flavor 
violation at an observable rate, unless the leptons would be localized 
closely towards the TeV-brane.

Constraints on the $K-\bar K$ mass splitting require the dimension-six 
operator $(ds)^2$ to be suppressed by $Q>5\times 10^{6}$ GeV. If it
contributes to CP violation, an even stronger suppression of 
$Q>5\times 10^{7}$ GeV is required. Using the quark locations of 
eq.~(\ref{qlocations}), we obtain $Q>7.2\times 10^{7}$ GeV for left-handed
and $Q>1.2\times 10^{8}$ GeV for right-handed states. For the operator
$(db)^2$ we find $Q>3.3\times 10^{6}$ GeV, which is above the corresponding
experimental bound of $2\times 10^{6}$ GeV. Other flavor violating operators like
$(cu)^2$ are also within their experimental bounds. The quark locations
of eq.~(\ref{qlocations}) which we obtained from the fermion mass pattern
automatically lead to the required suppression of flavor violating
non-renormalizable operators. This conforms the conclusions reached
in refs.~\cite{GP,HS2}.

Dimension-six operators contributing to proton decay are highly constrained
by experimental searches. For instance the operator  $q_1q_1q_2l_3$ has
to be suppressed by $Q>10^{15}$ GeV \cite{BD}. Taking again the 
quark locations of eq.~(\ref{qlocations}) and the lepton locations of
eq.~(\ref{llocations1}) we obtain $Q>5.6\times 10^7$ GeV. So some small
coupling of order $10^{-14}$ should multiply the non-renormalizable
operator to be consistent with experimental limits on proton decay. 
This might point to an additional symmetry, such as baryon or lepton 
number. A tiny coupling might also originate from non-perturbative effects
of gravity, especially if there is an extra dimension somewhat larger 
than $M_{\rm PL}$ \cite{KKLLS}. If the proton is stabilized by imposing
lepton number (or lepton parity in the case of Majorana neutrino masses),
baryon number violating processes like neutron-antineutron oscillations
could still occur \cite{CC01}. With the quark locations of eq.~(\ref{qlocations}) the
rate is, however, far below the experimental bound. 

In ref.~\cite{HS2}
somewhat different quark locations where used in order to maximally 
suppress proton decay by non-renormalizable interactions. Similar
conclusions were reached, even though the suppression scale $Q$
could be enhanced by an order of magnitude. Non-renormalizable 
interactions can be further suppressed if the fermions are shifted
closer towards the Planck-brane by allowing for larger 5D Yukawa
couplings. A smaller AdS curvature compared to the fundamental
Planck mass has the same effect. However, using these means to 
stabilize the proton would necessarily introduce new hierarchies in 
the model parameters.

\section{Conclusions}
In this paper we have studied aspects of flavor physics in the warped
SM. Gauge bosons and fermions are bulk fields, while the Higgs field
is confined to the TeV-brane. We have shown in detail how the fermion 
mass hierarchies and mixings can naturally be explained in a
geometrical way, without relying on hierarchical Yukawa couplings.
The observed fermion masses and mixings fix the relative positions
of the fermion fields in the extra dimension. Large mixings are
attributed to similar locations, and masses become small for
fermions localized towards the Planck-brane. 
Small neutrino masses can arise from a coupling to right-handed
neutrinos in the bulk or from dimension-five interactions.

Mixings between different KK levels induces deviations from the SM.
Electroweak fits require the KK scale to be at least 10 TeV.
KK mixings in the quark and weak gauge boson sectors lead to a 
non-unitary CKM matrix. However, deviations from unitarity are 
safely within the experimental bounds.  
 
Flavor violation by (KK) gauge boson exchange is an immediate 
consequence of our approach to the fermion mass problem. 
After transformation to fermion mass eigenstates, non-universal 
gauge couplings generate flavor changing neutral currents. Fixing
the fermion locations by the fermion masses and mixings,
we can predict the rates of flavor violations processes. Since
the light fermion flavors are localized towards the Planck-brane,
where non-universality is small, flavor violation is within experimental 
bounds even for a KK scale of 10 TeV. Some processes, such as 
muon-electron conversion, are in the reach of next generation experiments
and can provide valuable hints to the higher dimensional theory. 
This result is quite different from models with flat extra dimensions, 
where constraints on kaon mixing require the KK scale to be in the 
$10^3$ TeV range. Flavor violation in radiative decays is suppressed 
which is an important distinction to supersymmetric models. 

Bulk fermions also help to reduce the impact of non-renormalizable
operators. We have demonstrated that their contribution to flavor
violating processes is naturally below the experimental sensitivity,
even for order unity couplings. Dimension-six operators leading
to proton decay, however, cannot be sufficiently suppressed.
The required tiny couplings might point to some additional symmetry.  

Thus, the warped SM allows to generate the
fermion mass pattern from flavor-dependent locations without giving up the
solution to the gauge hierarchy problem or inducing
unacceptable rates for flavor violating processes.

\section*{Acknowledgements}
The author thanks David E.~Costa and Qaisar Shafi for valuable
discussions.

\end{document}